\documentclass[twoside,reqno]{HERON}
\usepackage{epsfig,cite,colordvi}
\usepackage{graphicx}
\usepackage{url}
\usepackage{amssymb,amsmath,amscd,epsf}
\usepackage{times}
\usepackage{makeidx}
\newcommand{\ds}{{\bf d}}

\begin{document}

\title{Systematic effects in the measurement of the negatively charged pion mass using laser spectroscopy of pionic helium atoms  }

\runningheads{Systematic effects in the measurement of the
negatively charge pion mass \ldots }{Boyan Obreshkov}

\begin{start}

\author{Boyan Obreshkov}{}

\index{Author, Boyan Obreshkov}

\address{Institute for Nuclear Research and Nuclear Energy\\ Bulgarian Academy of Sciences, Sofia 1784, Bulgaria }{}

\begin{Abstract}

The collision-induced shift and broadening of selected dipole
transition lines of pionic helium in gaseous helium at low
temperatures  up to $T=12$ K and pressure up to a few bar are
calculated within variable phase function approach. We predict
blue shift of the resonance frequencies of the $(n,l)=(16,15)
\rightarrow (16,14) $ and $(16,15) \rightarrow (17,14)$ unfavored
transitions and red shift for the favored transition $(17,16)
\rightarrow (16,15)$. The result may be helpful in reducing the
systematic error in proposed future experiments for determination
of the negatively charged pion mass from laser spectroscopy of
metastable pionic helium atoms.

\end{Abstract}
\end{start}

\section{Introduction}

Pionic atom spectroscopy permits to measure the charged pion mass.
Orbital energies of the system depend on the reduced mass of the
system. These energies can be calculated with high precision using
quantum electrodynamics (QED). Measuring transition frequencies,
not disturbed by strong interactions, allows to determine the
reduced mass of the system and hence the mass of the pion.

Two experiments at Paul Scherrer Institute (PSI) using X-ray
spectroscopy of pionic atoms gave most precise values for the
masses of the muon-type neutrino and of the charged pion. By
measuring the energy of the X-ray emitted in the 4f-3d transition
of pionic magnesium, the pion mass was measured to  3 parts per
million (ppm) accuracy \cite{JeckA,JeckB}. However the experiment
found a negative value for the square of the muon neutrino mass.
Re-analysis of the Mg-atom data including better model of electron
screening corrections, lead to a pion mass compatible with
positive mass squared of the muon-type neutrino.

Because of the complexity of the electronic structure of pionic
magnesium, more recently a method for laser spectroscopy of pionic
helium atoms has been proposed \cite{Hori2}. Pionic helium is a
three-body system composed of a helium nucleus, an electron in a
ground state and $\pi^-$ in highly excited nearly circular state
with principal and orbital quantum numbers $n \sim l+1$. These
states can de-excite via Auger transitions to lower lying states
which have large overlap with the helium nucleus and subsequently
undergo fast nuclear absorption for times less than a picosecond.
However long-lived $\pi^-$ were observed in bubble-chamber
experiments \cite{LL_pi}, to explain this anomaly, Condo
\cite{Condo} suggested that metastable atomic states of $\pi^-$
are formed
\begin{equation}
{\rm He}+ \pi^- \rightarrow [{\rm He}^+ \pi^-]_{nl} + e^-
\end{equation}
in which $\pi^-$ occupies highly excited Rydberg states with
principal quantum number $n \sim (m^*/m_e)^{1/2} \sim 16$, where
$m^*$ is the reduced mass of $\pi^-$ and the helium nucleus. For
nearly circular orbits $n \sim l+1$, the Auger decay rate is
strongly suppressed and because the radiative decay $\tau_{\rm
rad} \sim 1\ \mu s$ is also slow, the lifetime of the metastable
states is determined by the proper lifetime of $\pi^-$,
$\tau_{\pi^-} \sim 26$ ns. An indirect conformation of the Condo's
hypothesis has been obtained at TRIUMF \cite{piLiHe} in
experiments with $\pi^-$ stopped in liquid helium; it has been
found that about $2 \%$ of the pions retain a lifetime of 7 ns.

When comparing experimental transition frequencies to three-body
QED calculations of pionic helium, the $\pi^-$ mass  can be
determined with fractional precision better than 1 ppm
\cite{Hori1}. However systematic effects such as collision-induced
shift and broadening of the transition lines, the collision
quenching of the metastable pionic states, AC Stark shifts,
frequency chirp in the laser beam, can prevent the experiment from
achieving this high precision. Reliable theoretical calculation
for the density-dependent shift and width is needed for the
extrapolation of transition wavelengths at zero target density.

\section{Line shift and broadening calculations}

\subsection{Interatomic potentials}

The collisional shift and broadening of the laser stimulated
transition line in pionic helium are obtained in the impact
approximation of the binary collision theory of the spectral line
shape \cite{Lindholm,Baranger,Allard,anderson}. The interaction
energy between the pionic and ordinary helium atoms is obtained in
the Born-Oppenheimer approximation by separating adiabatically the
electronic from the nuclear degrees of freedom. The potential
energy surface (PES) for the description of the binary interaction
between an exotic helium with ordinary helium atom had been
evaluated with ab initio quantum chemistry methods
\cite{sapt,hfi14} for nearly 400 configurations of the two helium
nuclei and the pion. The nuclear configurations are parameterized
with the length $r$ of the vector joining the heavy particles in
the pionic atom, the length $R$ of the vector joining its
center-of-mass with the nucleus of the perturbing helium atom, and
the angle $\theta$ between them. Subsequently, the numerical
values of the PES at these 400 grid points were fitted with smooth
functions $V(r,R,\theta)$ and used in the calculation of the
collisional shift and dephasing rate. Because the collisional
quenching due to inelastic collisions at thermal energies is
unlikely to be relevant for the metastable states of the pionic
atom, we use central state-dependent potentials to describe the
elastic scattering
\begin{equation}
V_{nl}(R)= \frac{1}{2} \int dr |\chi_{nl}(r)|^2 \int d \theta \sin
\theta V(r,R,\theta),
\end{equation}
where $\chi_{nl}(r)$ are the unperturbed $\pi^-$ wave-functions.

\subsection{Impact approximation}

When the helium gas atoms are moving rapidly, the broadening and
shift the spectral lines arises from a series of binary encounters
between the pionic atom with ordinary helium atom. The impact
approximation is valid if the average time interval between
collisions is much larger than the duration of the collision
\cite{Baranger}. For an isolated spectral line produced by a
laser-stimulated dipole transition from an initial state $i=(n,l)$
to a final state $f=(n',l')$ of the pionic atom, the line shape
assumes a Lorentizan profile
\begin{equation}
I_{fi}(\omega)= \frac{|\ds_{fi}|^2}{\pi}
\frac{\Gamma_{fi}}{(\omega-\omega_{fi}+\Delta_{fi})^2
+\Gamma_{fi}^2},
\end{equation}
where $\ds_{fi}$ is the transition dipole moment,
$\omega_{fi}=E_f-E_i$ is the transition frequency and
$\Gamma_{fi}$ and $\Delta_{fi}$ are the collision-induced
broadening and shift, respectively. In the approximation of binary
collisions, both the shift $\Delta_{fi}=N \beta_{fi}$ and the
polarization dephasing rate $\Gamma_{fi}= N \alpha_{fi}$ are
linear functions of the gas density $N$. The slope of the
temperature-dependent collisional broadening and shift are
\begin{equation}
\alpha_{fi}(T)=\left \langle \frac{\pi}{M k} \sum_{L=0}^{\infty}
(2L+1) 2 \sin^2 \eta_{fi,L}(k) \right \rangle_T \label{barr1}
\end{equation}
and
\begin{equation}
\beta_{fi}(T)=-\left \langle \frac{\pi}{M k} \sum_{L=0}^{\infty}
(2L+1) \sin 2 \eta_{fi,L}(k) \right \rangle_T, \label{barr2}
\end{equation}
respectively, where $k =M v$ is the wave-number of relative
motion, $v$ is the impact velocity and $M$ is the reduced mass of
the collision system. Both $\alpha$ and $\beta$ and expressed in
terms of elastic scattering phase shifts
$\eta_{fi,L}(k)=\delta_{iL}(k)-\delta_{fL}(k)$ and $\langle F
\rangle_T$ is a thermal average
\begin{equation}
\langle F \rangle_T= 4 \pi \left( \frac{M}{2 k_B T} \right)^{3/2}
\int dv v^2 e^{-M v^2/ 2 k_B T} F(v)
\end{equation}
over the Maxwell velocity distribution at temperature $T $  for
the pionic helium-helium system and $k_B$ is the Boltzmann
constant. The partial wave phases $\delta_L(k)=\delta_L(k,R
\rightarrow \infty)$ are obtained from the asymptotic solution of
the variable phase equation \cite{pam}
\begin{equation}
\frac{d}{dR} \delta_{nlL}(k,R)=-\frac{2M V_{nl}(R)}{k}  [\cos
\delta_{nlL}(k,R) j_L(kR)-\sin \delta_{nlL}(k,R)n_L(kR)]^2,
\end{equation}
subject to the boundary condition $\delta_{nlL}(k,0)=0$, and
$\{j_L(z),n_L(z)\}$ are the Riccati-Bessel functions.

\section{Numerical results and discussion}

Tab. \ref{tab:tr1} and Tab. \ref{tab:tr2} present values for the
scattering phase shifts $\eta_{fi,L}(k)$ for the "favored"
transition $(17,16) \rightarrow (16,15)$ and for the "unfavored"
one $(16,15) \rightarrow (16,14)$, respectively. For the unfavored
transition in Tab. \ref{tab:tr1}, scattering phases are
appreciably less than 1 radian over the whole range of wave
numbers. In this regime of weak collisions, the phase shifts are
added linearly and contribute to the line shift $\Delta$, but have
little effect on the broadening $\Gamma$. Since all phases are
negative, the transition frequency undergoes a blue shift. For $k
< 0.5$, the scattering of $s,p$ and $d$-waves gives dominant
contribution to the dipole transition lineshape, contributions of
partial waves with $L \ge 3$ are suppressed due to large centrifugal
barrier. Larger number of partial waves is required to converge
$\Delta$ and $\Gamma$ for $k > 0.5$. The elastic scattering phase
shifts associated with the favored transition are shown in Tab.
\ref{tab:tr2}, scattering phases are positive and have appreciable
values resulting in red-shift and substantial broadening of the
transition line. The principal contribution to the shift and
broadening is due $s$- and $p$-wave scattering, the $d$-wave
scattering is less pronounced. Since the $s$-wave phase shift
$\eta_0 \rightarrow \pi$ near $k \rightarrow 0$, the scattering
potential in the initial state $V_{17,16}(R)$ supports a single
bound state. Because the $s$-wave scattering is dominant towards
threshold, this bound state affects dramatically the transition
line shape at very low speeds with $k < 0.1$ as the center
frequency undergoes a blue shift when $\eta_0
> \pi/2$. However the thermally-averaged shift and width are
weakly affected by the low-velocity tail in the Maxwell
distribution.

Table~\ref{tab:tmp}  presents numerical results on the temperature
dependence of the slopes of the collisional shift and broadening,
$\beta_{fi}(T)$ and $\alpha_{fi}(T)$, of the two transition lines
of known experimental interest \cite{private}. The temperature
dependence of the line profile is relatively weak in gaseous
helium. At low perturber density $N=10^{21}$ cm$^{-3}$, the
resonance frequency of the unfavored transitions $(n,l)=(16,15)
\rightarrow (16,14)$  and $(16,15) \rightarrow (17,14)$ are blue
shifted with $\Delta \approx 2.5$ GHz and $\Delta=18$ GHz,
respectively. For the favored transition $(17,16) \rightarrow
(16,15)$ the line center undergoes a red-shift with $\Delta
\approx -8$ GHz. The large collisional broadening of the
transition line $(16,15) \rightarrow (17,14)$ $\Gamma=7.7$ GHz
would not allow the transition wavelength to be determined to a
fractional precision better than 1 ppm.  Because the collisional
broadening  of the transition line $(16,15) \rightarrow (16,14)$
is $\Gamma=0.1$ GHz, this unfavored transition is suitable for
spectroscopic measurements in pionic helium.

\begin{figure}
\begin{center}
\includegraphics{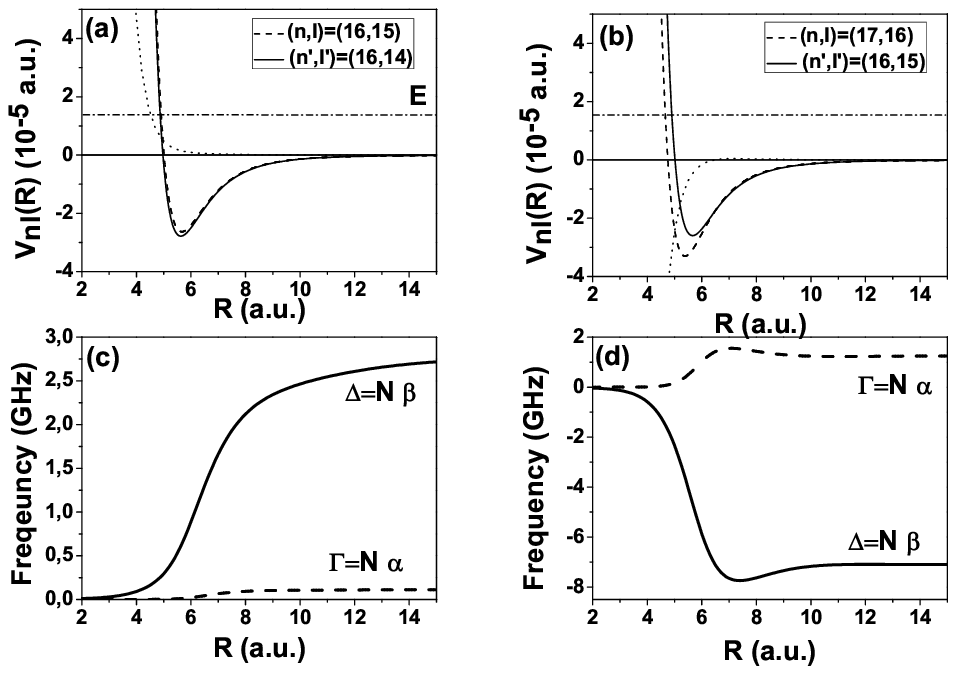}[htb]
\caption{(a) and (b) Potential energy curves $V_{nl}(R)$ for the
elastic scattering of pionic helium atoms by an ordinary helium
atom prior to (dashed line)  and after (solid line) the absorption
of a photon in laser-stimulated dipole transitions $(16,15)
\rightarrow (16,14)$ and $(17,16) \rightarrow (16,15)$,
respectively. The potential energy difference $\Delta
V=V_{nl}-V_{n'l'}$ is given by a dotted line, and the kinetic
energy of relative motion is indicated by the dashed-dotted line.
(c) and (d) Variable line shift and broadening (in GHz)
corresponding to the potential energy curves in (a) and (b).  }
\label{fig:line_Rd}
\end{center}
\end{figure}

To further analyze the effect of the interaction energy
$V(r,R,\theta)$ in the ${\rm He}^+ \pi^-$-${\rm He}$ collision
system, in Fig.\ref{fig:line_Rd}(a-b) we plot the effective state
dependent potentials for the transitions $(17,16) \rightarrow
(16,15)$ and $(16,15) \rightarrow (16,14)$, respectively, together
with the potential energy differences $\Delta V=V_i-V_f$. A
general property of the state-dependent potentials is that they
are short-ranged, exhibit a hard repulsive part for $R < 5$ a.u.
and display a potential minimum located near $R \approx 6$ a.u..
In the variable phase-function approach, Fig.
\ref{fig:line_Rd}(c-d) represent the corresponding
position-dependent line shift $\Delta(R)=N \beta(R)$ and
broadening $\Gamma(R)=N \alpha(R)$ functions at thermal collision
energy with $T=6$ K. The variable line-shift $\Delta(R)$ and width
 $\Gamma(R)$ functions are defined through Eq.(\ref{barr1}) and
Eq.(\ref{barr2}) in terms of the local phase shift
$\eta_{fi}(k,R)=\delta_i(k,R)-\delta_f(k,R)$. For the unfavored
transition, the phase functions  involve primary weak distant
collisions with $R > 5$ a.u. The broadening function saturates
rapidly until $R=8$ a.u., while the shift function saturates much
more slowly due to the contributions of higher partial waves with
$L>2$ (large impact parameter). The contribution of close-range
binary encounters with $L=0,1$ is weakened, because of smaller
statistical weight $(2L+1)$. In contrast, the line shape of the
favored transition is primary determined by stronger collisions
involving $s$- and $p$-wave scattering. Essential part of the line
shift (and width) comes from the classically forbidden region for
the relative motion with $4 < R \le 5$ a.u. the shift function rises
steeply in the classically allowed part of the scattering
potentials $R > 5$ a.u., attains maximum near $R \approx 7$ a.u.,
then slightly falls off and saturates in the asymptotic region with
$R > 9$ a.u. Thus for this favored transition, the principal part
of the line shift and width at thermal collision energies is due
to short-range binary encounters, in this case the effect of the
long-range van-der-Waals tail $V(R) \sim C_6/R^6$ can be treated
as a weak perturbation.

\begin{table}
\caption{ Phase analysis of the collisional shift and broadening
of dipole transition line shape $(n,l)=(16,15) \rightarrow
(n'l')=(16,14)$ in pionic helium interacting with gaseous helium
at temperature $T=6$ K.. Partial phase shifts
$\eta_{fi}(k),L=0,\ldots,8$ in radians, $k$ is the wave-number of
relative motion. } \label{tab:tr1}

\begin{tabular}{|c|ccccccccc|}
\hline
 \multicolumn{1}{|r|}{$L=$} &
 \multicolumn{1}{c}{0} &
 \multicolumn{1}{c}{1} &
 \multicolumn{1}{c}{2} &
 \multicolumn{1}{c}{3} &
 \multicolumn{1}{c}{4} &
 \multicolumn{1}{c}{5} &
 \multicolumn{1}{c}{6} &
 \multicolumn{1}{c}{7} &
 \multicolumn{1}{c|}{8} \\
 \multicolumn{1}{|l|}{$k$, a.u.}
 & \multicolumn{9}{c|}{$\eta_{fi}(k)$, rad}\\
 \hline
 0.127 &
 -0.126  &  -0.047  &  -0.001 &   0.000  &  0.000  &
 0.000  &  -0.000 &   0.000 &   0.000  \\
 \hline
 0.310
 & -0.063  &  -0.067  &  -0.041 &   -0.007  &  -0.001  &
 0.000  &  0.000  &  0.000  &  0.000  \\
 \hline
 0.538
 & -0.047  & -0.048  &  -0.047 &   -0.036  &  -0.014   &
 -0.004  &  -0.001  &  0.000  &  0.000  \\
 \hline
 0.805
 & -0.044  & -0.043  & -0.041 &   -0.038  &  -0.032  &
 -0.019  &  -0.009  &  -0.004  &  -0.002  \\
  \hline
 1.131
 & -0.046  & -0.045  & -0.043 &  -0.040  & -0.035  &
 -0.030  &  -0.022  &  -0.013  &  -0.007  \\
 \hline
 \end{tabular}
 \end{table}

\begin{table}
\caption{ Phase analysis of the collisional shift and broadening
of dipole transition line shape $(n,l)=(17,16) \rightarrow
(n'l')=(16,15)$ in pionic helium interacting with gaseous helium
at temperature $T=6$ K. Partial phase shifts
$\eta_{fi}(k),L=0,\ldots,8$ in radians, $k$ is the wave-number of
relative motion. } \label{tab:tr2}

\begin{tabular}{|c|ccccccccc|}
\hline
 \multicolumn{1}{|r|}{$L=$} &
 \multicolumn{1}{c}{0} &
 \multicolumn{1}{c}{1} &
 \multicolumn{1}{c}{2} &
 \multicolumn{1}{c}{3} &
 \multicolumn{1}{c}{4} &
 \multicolumn{1}{c}{5} &
 \multicolumn{1}{c}{6} &
 \multicolumn{1}{c}{7} &
 \multicolumn{1}{c|}{8} \\
 \multicolumn{1}{|l|}{$k$, a.u.}
 & \multicolumn{9}{c|}{$\eta_{fi}(k)$, rad}\\
 \hline
 0.127 &
 0.382  &  0.115  &  0.000 &   0.000  &  0.000  &
 0.000  &  0.000 &   0.000 &   0.000  \\
 \hline
 0.310
 & 0.243  &  0.226  &  0.083 &   0.002  &  -0.002  &
 0.000  &  0.000  &  0.000  &  0.000  \\
 \hline
 0.538
 & 0.230  & 0.219  &  0.183 &   0.087  &  0.009   &
 0.000  &  0.000  &  0.000  &  0.000  \\
 \hline
 0.805
 & 0.248  & 0.238  & 0.217 &   0.177  &  0.103  &
 0.025  &  0.006  &  -0.001  &  0.000  \\
  \hline
 1.131
 & 0.280  & 0.273  & 0.257 &  0.231  & 0.192  &
 0.133  &  0.061  &  0.013  &  0.000  \\
 \hline
 \end{tabular}
 \end{table}

\begin{table}
\caption{Slope of the density shift and broadening $\beta(
\alpha)$ for selected transitions in pionic helium and
temperatures in the range $4-12$ K, in units of $10^{-21}$
GHz.cm$^3$ } \label{tab:tmp}
\begin{center}
\begin{tabular}{|lrrrrrr|}
\hline \multicolumn{1}{|l|}{Transition} & \multicolumn{1}{c}{$T$
(K)} & \multicolumn{1}{c}{4} & \multicolumn{1}{c}{6} &
\multicolumn{1}{c}{8} & \multicolumn{1}{c}{10} &
\multicolumn{1}{c|}{12} \\
%\\
\hline
$(16,15) \rightarrow (16,14)$ & & 2.97(0.16) & 2.93(0.14) & 2.92(0.13) & 2.93(0.12) & 2.94(0.11) \\
$(16,15) \rightarrow (17,14)$ & & 17.56(7.73) & 17.97(7.19) & 17.95(6.98) & 17.94(6.68) & 18.07(6.42) \\
$(17,16) \rightarrow (16,15)$ & & -7.45(1.55) & -7.55(1.41) & -7.78(1.43) & -8.00(1.47) & -8.23(1.52) \\
\hline
\end{tabular}
\end{center}
\end{table}

 \section{Conclusion}

We calculated the density shift and broadening of selected dipole
transition lines in pionic helium in gaseous helium. At thermal
collision energies, we find blueshift of the line center of the
unfavored transitions $(n,l)=(16,15) \rightarrow (16,14)$ and
$(n,l)=(16,15) \rightarrow (17,14)$, the transition frequency is
red-shifted for a favored transition $(17,16) \rightarrow
(16,15)$.  The negligible collisional broadening ($\Gamma=0.1$
GHz) of the resonance transition $(n,l)=(16,15) \rightarrow
(16,14)$ makes it suitable candidate for precision spectroscopy of
pionic helium atoms. The theoretical result may be helpful in the
extrapolation of the transition wavelengths in pionic helium to
zero density of the perturbing helium gas.

\end{document}